\documentclass[pra,amsmath,aps,twocolumn,superscriptaddress]{revtex4-1}
\usepackage{amsmath,mathrsfs,amsbsy,amssymb,lmodern,graphicx,bm,amsthm,amsfonts}
\usepackage{units}
\usepackage{bbm}
\usepackage{multirow,color}
\usepackage{subfigure}
\usepackage{xcolor}
\usepackage{lineno}

\begin{document}

\title{Experimentally realized local-global tradeoff in quantum resources}
\author{Yue Dai}
\thanks{These two authors contributed equally}
\affiliation{School of Physical Science and Technology, Ningbo University, Ningbo, 315211, China}
\affiliation{School of Optical and Electronic Information, Suzhou City University, Suzhou, 215104, China}
\author{Zhiyu Wang}
\thanks{These two authors contributed equally}
\affiliation{School of Physical Science and Technology, Ningbo University, Ningbo, 315211, China}
\author{Xinzhi Zhao}
\affiliation{School of Physical Science and Technology, Ningbo University, Ningbo, 315211, China}
\affiliation{College of Physics, Nanjing University of Aeronautics and Astronautics, Nanjing, 211106, China}
\author{Xinglei Yu}
\affiliation{School of Physical Science and Technology, Ningbo University, Ningbo, 315211, China}
\affiliation{Department of Mechanical and Automation Engineering, The Chinese University of Hong Kong, Hong Kong}
\author{Chengjie Zhang}
\email{chengjie.zhang@gmail.com}
\affiliation{School of Physical Science and Technology, Ningbo University, Ningbo, 315211, China}

\begin{abstract}
The interconversion among different quantum resources has attracted considerable attention, exemplified by the tradeoff between coherence and entanglement. However, the relationship between local and global quantum resources remains largely unexplored. Here, we propose and experimentally demonstrate a local global tradeoff in quantum resources, quantified respectively by the local maximum coherence and the quantum mutual information. Our experiments show that the sum of the local maximum coherence and the quantum mutual information is bounded by a definitive upper limit. These results not only establish a fundamental tradeoff inequality, but also provide direct experimental evidence of the complementarity between local and global quantum resources. Moreover, our findings reveal an important constraint in quantum networks: the local quantum resources available at individual nodes and the global quantum resources of the network are mutually exclusive, implying that enhancing one necessarily limits the other.
\end{abstract}

\maketitle



\section{Introduction}
In quantum physics, phenomena that markedly depart from classical intuition have long been a central focus of research, including quantum coherence~\cite{coh} and quantum entanglement~\cite{ent}. These phenomena can be broadly classified into two categories: those characterizing intrinsic quantum effects within a single system, such as quantum coherence, and those describing correlations among multiple quantum systems, such as quantum entanglement and quantum mutual information (QMI)~\cite{qcqi}. These quantum features underpin a wide range of applications, including quantum key distribution~\cite{qkd1,qkd2,qkd3,qkd4,qkd5}, quantum teleportation~\cite{qt1,qt2,qt3,qt4,qt5}, and quantum computing~\cite{qc1,qc2,qc3,qc4,qc5}. As quantum technologies continue to advance, the precise quantification of these quantum phenomena has become increasingly important~\cite{qrt}.
Notably, different quantum resources share structural similarities in their quantification, providing a unified perspective for investigating and comparing diverse quantum phenomena. 

Quantum coherence is one of the most fundamental properties that distinguishes quantum physics from classical physics. The preservation of quantum coherence often signifies the stability of quantum systems. In quantum information applications, qubits serve as basic operational units, and the reliability of a qubit is typically determined by the coherence of its physical carrier~\cite{qcqi}. Several measures of quantum coherence exist, including the $l_1$-norm of coherence and the relative entropy of coherence~\cite{coh}. Although quantum coherence is typically defined in a basis-dependent manner, to enable clearer experimental observation of the tradeoff relationship, we adopt the maximum coherence for the relative entropy of coherence, which is basis-independent~\cite{mco1, mco2}.

\begin{figure}[tp]
\centering
\includegraphics[scale=0.6,clip=true,trim=20 20 20 20]{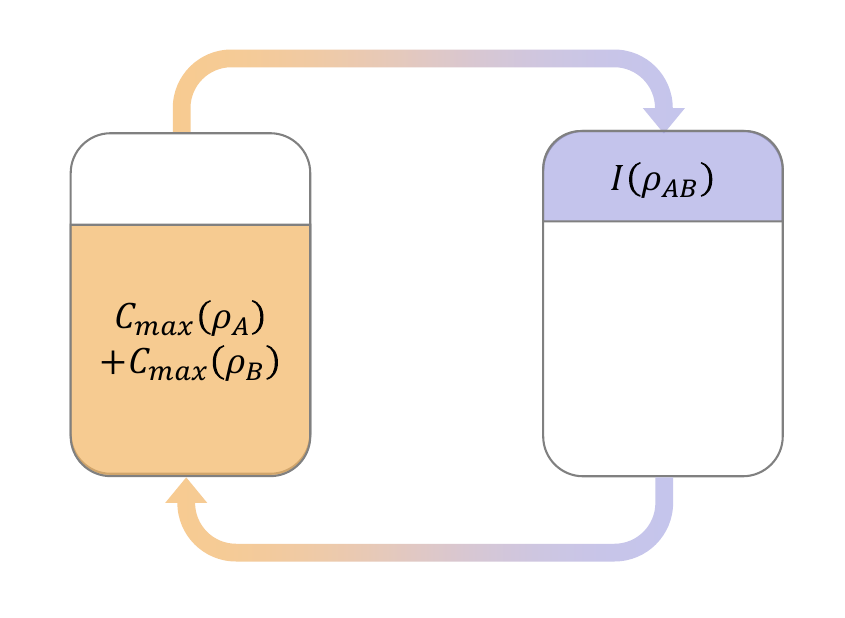}
\caption{The local-global tradeoff in quantum resources. The sum of local maximum coherence $C_{\mathrm{max}}(\rho_A)+C_{\mathrm{max}}(\rho_B)$ denotes the local quantum resource (left), and quantum mutual information $I(\rho_{AB})$ is the global resource (right). Our experiments demonstrate that the above two resources can be transformed into each other in a two-qubit system.  The orange indicates local resource, i.e., the local maximum coherence. The lavender indicates global resource ---  quantum mutual information. Arrows illustrate the conversion between the two resources.}
\label{trade}
\end{figure}

\begin{figure*}[tp]
\centering
\includegraphics[scale=0.13]{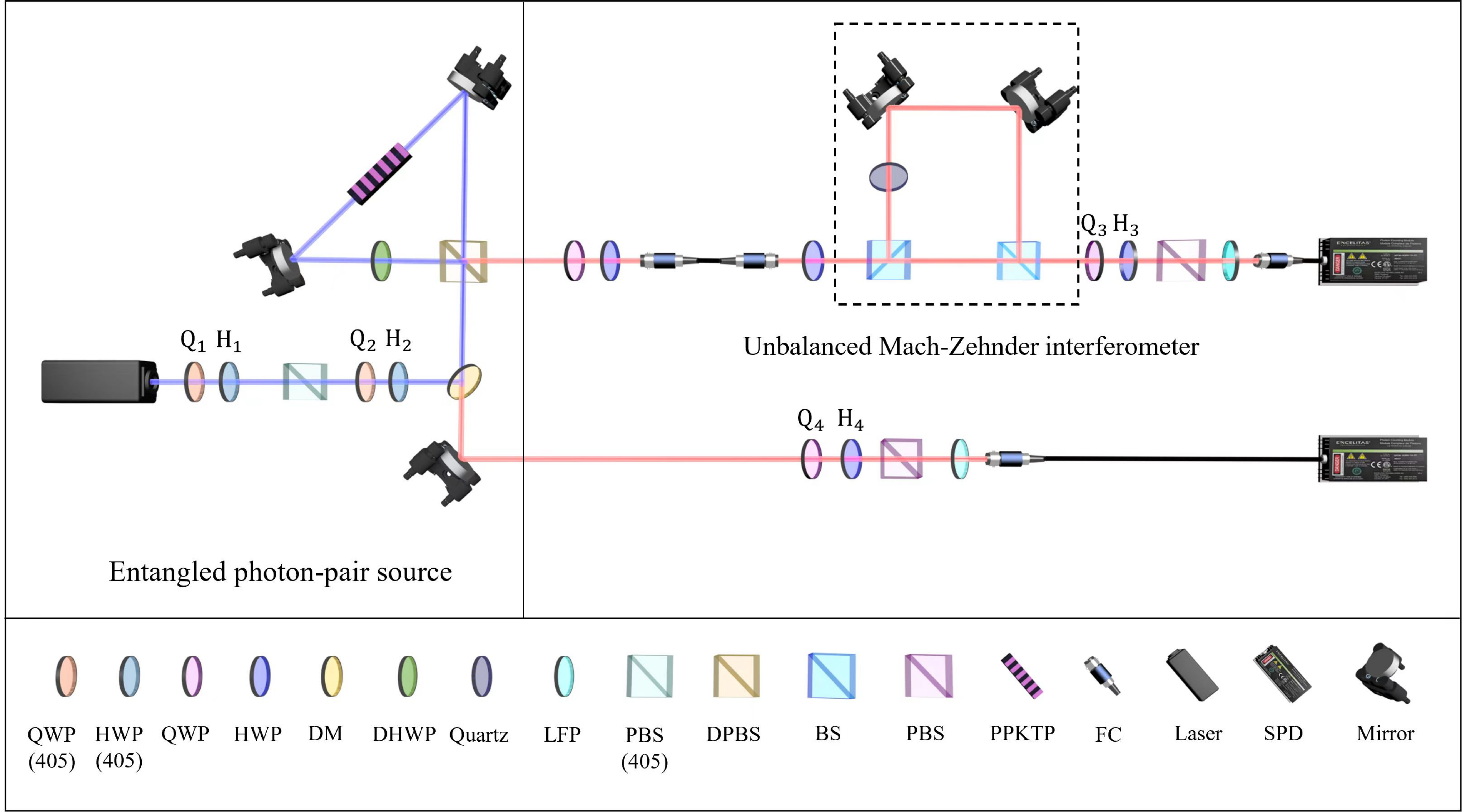}
\caption{The experimental setup. The first PBS polarize the photons from the cw laser. Q$2$ and H$2$ used to adjust the polarization direction of the light source.  The polarization-entangled photon pairs are generated by the spontaneous parametric down-conversion
process of the PPKTP. The PPKTP used in the experiment only responds to horizontally polarized light, so we set up a DHWP rotated 45¡ã on the originally vertically polarized light path of the PPKTP side. An unbalanced Mach-Zehnder interferometer in the dashed pane, which separates the photon into long and short paths, is inserted into up part to prepare mixed states. We inserted different numbers of BSs into two paths respectively to change the ratio of the number of photons in the two channels, so that we can obtain different mixed states. If the experiment requires a pure state, we only need to remove the part in the dashed pane. In each part, HWP~(H$3$, H$4$), QWP~(Q$3$, Q$4$), and PBS are set for state tomography. The photons are collected by two single-photon counting modules and identified
by the coincidence counter.}
\label{exp}
\end{figure*}

On the other hand, mutual information quantifies the total amount of information shared between different systems. In the quantum regime, quantum mutual information (QMI) is defined using the von Neumann entropy~\cite{qcqi}, in contrast to the Shannon entropy employed in classical information theory. Quantum mutual information captures both classical and quantum correlations, where the latter include not only quantum entanglement but also more general forms of nonclassical correlations. Owing to its comprehensive characterization of correlations, quantum mutual information has played an important role in several emerging research areas, such as quantum biology~\cite{qmib1,qmib2} and quantum machine learning~\cite{qml1,qml2}.
Tradeoff relations among different quantum resources have long been a central topic in quantum information theory~\cite{trade1,trade2,trade3,trade4,trade5,trade6,trade7,trade8,trade9,trade10,trade11,trade12,trade13,trade14,trade15,trade16}. While substantial effort has been devoted to understanding the tradeoff between quantum coherence and entanglement~\cite{cama1,cama2}, the tradeoff between local and global quantum resources remains comparatively unexplored.

In this work, we present theoretical results and experimental demonstrations of the tradeoff between local and global quantum resources, as illustrated in Fig.~\ref{trade}. We construct bipartite quantum states using entangled photon pairs, where two qubits are encoded in the polarization states of the photons. Quantum state tomography is employed to reconstruct the density matrices of these quantum states, allowing for direct experimental observation of the tradeoff for both pure and mixed states. Our experiments present direct experimental evidence of the tradeoff between local and global quantum resources in bipartite systems, which are theoretically extendable to multipartite scenarios.

\section{Local-global tradeoff in quantum resources}
For a quantum state $\rho$, the relative entropy of coherence is $C(\rho)= S(\rho_{\mathrm{diag}})-S(\rho)$ \cite{coh}, where $S(\rho)$ is the von Neumann entropy of $\rho$, and $\rho_{\mathrm{diag}}$ denotes the quantum state removing all off-diagonal elements of $\rho$ under the chosen basis $\{|i\rangle\}$. This quantum coherence measure is basis-dependent. In the following, we use a basis-independent coherence measure, i.e., maximum coherence~\cite{mco1,mco2}, instead of the relative entropy of coherence. The maximum coherence for  a quantum state $\rho$ is defined as $C_{\max}(\rho):=\big[\max_{\{|i\rangle\}} S(\rho_{\mathrm{diag}})\big]-S(\rho)$ \cite{mco1}. This implies that for any state $\rho$, an optimal basis can be found to maximize $S(\rho_{\mathrm{diag}})$, resulting in $C_{\max}(\rho) = \log_2 d - S(\rho)$~\cite{mco1}, where  $d$ is  the dimension of the Hilbert space. Maximum coherence is considered to be a more intrinsic definition of coherence after optimization.

Consider a bipartite state $\rho_{AB}$ shown in Fig.~\ref{trade}, the sum of local maximum coherence $C_{\mathrm{max}}(\rho_A)+C_{\mathrm{max}}(\rho_B)$ naturally indicates maximal local quantum resource of subsystems $A$ and $B$.
Let us present the tradeoff relation of local and global quantum resources, which are quantified by local maximum coherence and quantum mutual information, respectively. For any finite-dimensional bipartite state $\rho_{AB}$, the local maximum coherences $C_{\max}(\rho_A)$, $C_{\max}(\rho_B)$, and the QMI $I(\rho_{AB})$ satisfy the following tradeoff inequality~\cite{SM}
\begin{equation} \label{trade}
C_{\max}(\rho_{A})+C_{\max}(\rho_{B})+I(\rho_{AB})\leq \log_2(d_A d_B),
\end{equation}
where $d_A$ ($d_B$) is the dimension of subsystem $A$ ($B$), $I(\rho_{AB})=S(\rho_{A})+S(\rho_{B})-S(\rho_{AB})$ is quantum mutual information,  and $\rho_{A}$ ($\rho_{B}$) is reduced density matrix obtained by tracing out subsystem $B$ ($A$) from  $\rho_{AB}$. The equality in Eq. (\ref{trade}) holds if and only if $\rho_{AB}$ is a pure state. A detailed proof of inequality (\ref{trade}) can be found in \cite{SM}. Obviously, $C_{\max}(\rho_A)$ and $C_{\max}(\rho_B)$ represent the local resources of subsystem $A$ and $B$, respectively, while $I(\rho_{AB})$ represents the global quantum resource between the two subsystems, which encompasses both quantum and classical correlations. This formula indicates that in the bipartite system, there is a tradeoff between local and global quantum resources. The upper bound of the inequality corresponds to the maximum entropy of the bipartite system, which can also be understood as the maximum amount of available resources that the system may contain.

We can interpret this inequality from the perspective of resource allocation. The various available resources, local or global, will each occupy a portion of the total resources. The sum of all available resources in a system cannot exceed the total resources. For pure states, the sum of the local maximum coherence and QMI can always reach the maximum value $\log_2{(d_A d_B)}$. Thus, the larger the local resource, the smaller the global resource between two subsystems and vice versa.

To experimentally verify inequality (\ref{trade}), we generate two types of bipartite quantum states: a pure state and a mixed state. For the pure state, we directly observe the tradeoff between local maximum coherence and QMI. In contrast, for the mixed state, we observe that the sum of the local maximum coherence and QMI remains strictly below a fixed upper bound.

We begin with the pure-state scenario. The state is prepared as
\begin{equation} \label{rs1}
\left|\varPsi_{q} \right>=\sqrt{q}\left|01\right>-\sqrt{1-q}\left|10\right>,
\end{equation}
where the parameter
$q$ is experimentally tunable. The local maximum coherence and QMI for this state,
$\left|\varPsi_{q} \right>$, are given by  $C_{\max}(\rho_{A})+C_{\max}(\rho_{B})=2+2q\log_2q+2(1-q)\log_2(1-q)$ and $I(\rho_{AB})=-2q\log_2q-2(1-q)\log_2(1-q)$, respectively. Summing these quantities yields a constant value of 2, which serves as a fixed upper bound in this case.

Next, we consider the mixed-state scenario. The mixed state is constructed as
\begin{equation} \label{ms1}
\rho_m=p\left|\varPsi_+\right>\left<\varPsi_+\right|+(1-p)\left|\varPsi_-\right>\left<\varPsi_-\right|,
\end{equation}
where $\left|\varPsi_{\pm}\right>=~(\left| 01\right>\pm\left| 10\right>)/{\sqrt{2}}$ are two maximally entangled Bell states, and $p$ is a controllable parameter. According to Eq.~(\ref{trade}), the sum of the local maximum coherence and QMI for $\rho_m$ is $2+p\log_2p+(1-p)\log_2(1-p)\le 2$, demonstrating that the total is always bounded above by  the fixed value 2.

\begin{figure}[tp]
\centering
\includegraphics[scale=0.35]{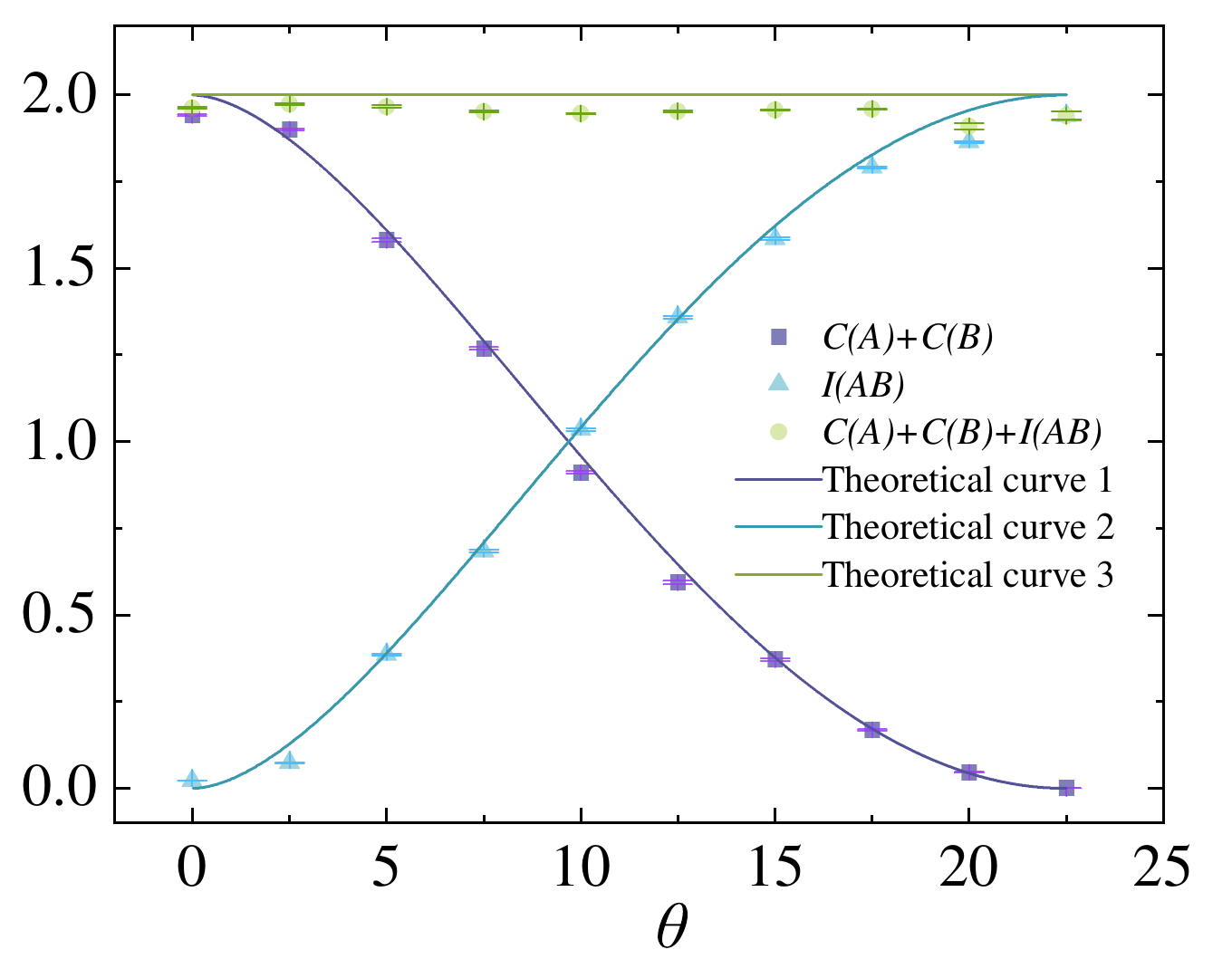}
\caption{The result of pure states. Here we calculate three sets of data. The light green solid line is the theoretical value of the upper bound of the inequality, while the round points are experimental values. The light purple solid line is the theoretical value of the local maximum coherence, while the square points are experimental values. The light blue solid line is the theoretical value of the QMI, while the triangle points are experimental values. As the light green solid line rises, the light purple solid line falls, This shows there is a tradeoff between local maximum coherence and the QMI.}
\label{puredata}
\end{figure}

\section{Experimental setup}
In our experiment, we can achieve both pure and mixed states by designing only one optical quantum system. The experimental setup is shown in Fig.~\ref{exp}. The degrees of freedom of polarization are used to represent the two bases of qubits, i.e., $\left|0 \right>\rightarrow\left|H \right>$ and $\left|1 \right>\rightarrow\left|V \right>$, which denote the horizontal and vertical polarization states of a photon, respectively.

We used  a continuous wave laser with a wavelength of $405$ nm. The quarter-wave plate~(QWP) $\mathrm{Q}_2$ is set to $0^{\circ}$ and the half-wave plate ~(HWP) $\mathrm{H}_2$ is set to $-22.5^{\circ}$, 
so that the input state is $\left|\psi\right> =~(\left| H\right> -\left| V\right>)/\sqrt{2}$. In this way, we can ensure that the number of photons entering the periodically poled KTiOPO$_4$ ~(PPKTP) crystal is approximately the same on both sides after passing through a polarizing beam splitter ~(PBS). A two-qubit state of approximately $\left|\varPsi_-\right>$ is produced through a spontaneous parametric down-conversion~(SPDC) process \cite{SPDC} in the PPKTP crystal.  An unbalanced Mach-Zehnder interferometer~\cite{unbalanced}  in the dashed pane, which separates the photon into long and short paths, is inserted into the upper part to prepare mixed states. The time difference between long and short paths is much larger than the coherent time of the photon. As a result, the generated state becomes a mixed state $\rho_m=p\left|\varPsi_+\right>\left<\varPsi_+\right|+~(1-p)\left|\varPsi_-\right>\left<\varPsi_-\right|$, where $p$ can be controlled in the dashed pane. Finally, the evolved state is reconstructed by tomography. QWPs, HWPs, and PBSs are used to set the $16$ measurement bases.

\section{Experimental results}
The experimental results of the pure state are shown in Fig. \ref{puredata}. During this experiment, we removed the dashed pane in the upper path. By changing the angle of H$2$, the number of photons on both sides of the PPKTP crystal is in a different ratio to achieve the pure state $\left|\varPsi_{q} \right>=\sqrt{q}\left|HV\right>-\sqrt{1-q}\left|VH\right>$, where $q=\cos^2{2\theta}$ and $1-q=\sin^2{2\theta}$. Here $\theta$ is the angle of rotation in the counterclockwise direction of H$2$. The lines in the figure represent the results of the theoretical calculations and the dots represent the experimental results. The light green represents the sum of the local maximum coherence and the QMI, which is the upper bound of the inequality. Light blue represents the QMI representing the global correlation. Light purple is the sum of the maximum coherence of the two sub-states. See the supplementary material for the data source~\cite{SM}. The experimental data represented by the points are very close to the theoretical calculation data represented by the lines, indicating that our experimental data are credible. In the previous analysis, we pointed out that in the case of pure states, the sum of local maximum coherence and QMI always reaches the upper bound. Therefore, we can clearly observe in the figure that there is a tradeoff relation between the QMI represented by light blue and the local maximum coherence represented by light purple.

\begin{figure}[tp]
\centering
\includegraphics[scale=0.35]{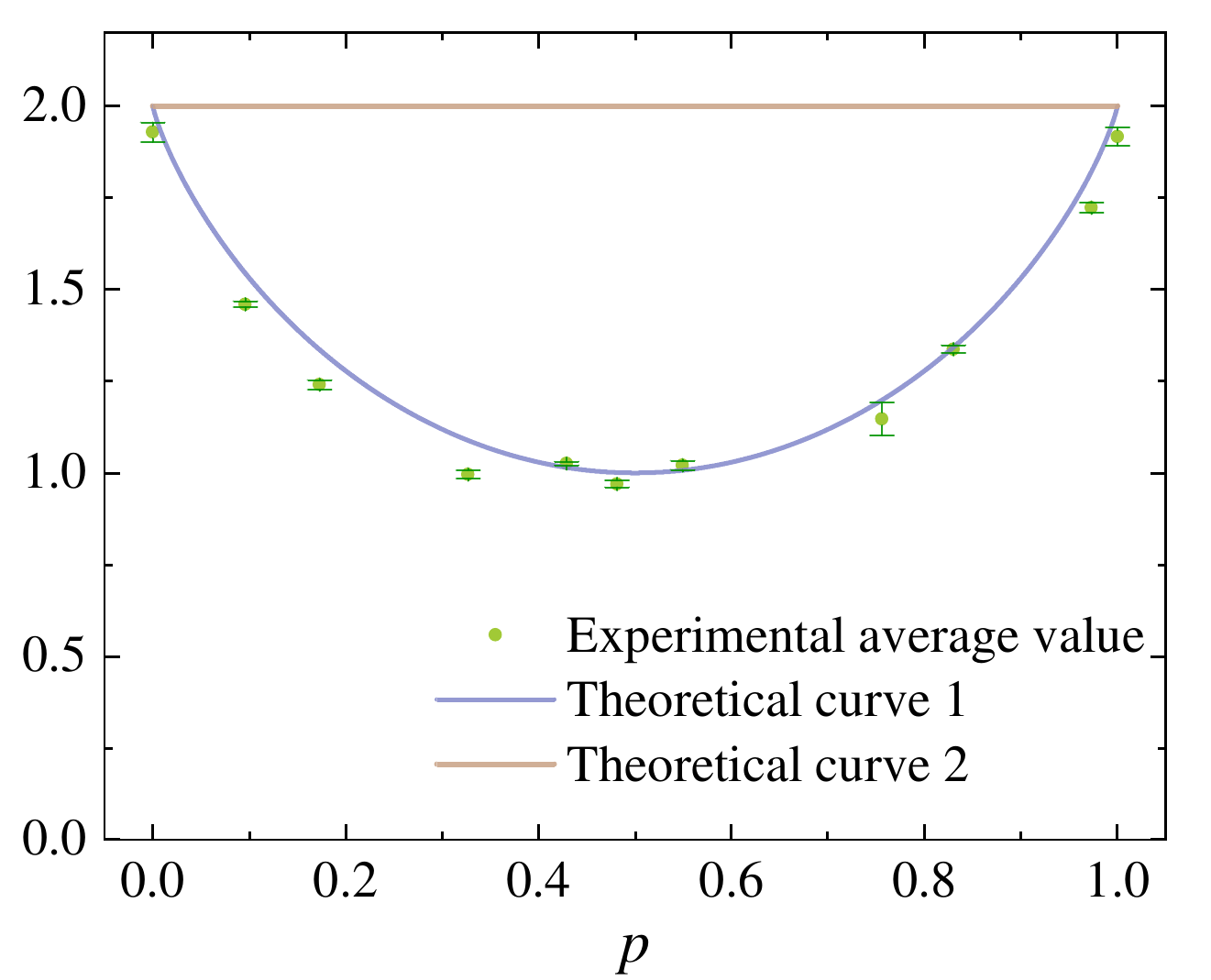}
\caption{The result of mixed states. In this experiment we just want to know whether the upper bound on the inequality still holds, so we only calculate one set of experimental data. The light green points represent the experimental values of the sum of local maximum coherence and the QMI, while the blue solid line is the theoretical value. The brown solid line is the upper bound of the inequality. The green points are all below the brown solid line.}
\label{mixdata}
\end{figure}


Let us analyze the experimental results of the mixed state in Fig. \ref{mixdata}. The solid brown line represents the upper bound of the left-hand side of Eq.~(\ref{trade}) with the mixed state $\rho_m$. The blue line is the result of the theoretical calculation of $2+p\log_2p+(1-p)\log_2(1-p)$. The light green dots represent the experimental results with different parameters $p$. The value $p$ is obtained from the ratio of the number of photons in the long and short optical paths within the dashed pane. We inserted different numbers of beam splitter~(BS) in the long and short paths, respectively, in the dashed pane to change the ratio of the number of photons in the two paths to achieve different $p$. The experimental data are also found in the supplemental material~\cite{SM}. From the figure, we can see that the experimental results are relatively close to the theoretical calculation results, which shows that our experimental results are credible. In the case of a mixed state, the sum of the local maximum coherence and quantum mutual information of the two substates often cannot reach the upper bound. In the figure, we can also see that the blue line is always below the brown line except for the two endpoints. This set of experimental data mainly shows the reliability of the upper bound of the tradeoff inequality.

It is worth noting that although we did not observe a tradeoff relation in the mixed-state experiment, this does not mean that such a relation does not exist in the mixed state. In mixed states, the sum of local coherence and QMI also have tight upper bounds. The sum of different types of resource in a state must not exceed the maximum available resources that the state may contain. Compared with pure states, part of the resource space of the mixed state will be occupied by entropy, and entropy cannot be utilized. The total resource space minus the entropy part is the maximum available resource. Different mixed states have different maximum available resources, which leads to different tight upper bounds. This is why we cannot easily observe the tradeoff relation in mixed states.

\section{Discussions and conclusions}
We propose a novel tradeoff relation between local quantum resources and global correlation. Our approach differs from previous studies in two key aspects. First, we replace coherence with maximum coherence. The measure of maximum coherence is basis independent, making it a more intrinsically local resource. Second, we use QMI to represent global correlation, highlighting that local quantum resources are not only tradeoff related to quantum correlations but also to classical correlations. In quantum networks, classical correlations can also serve as a resource~\cite{Hawkins2025}. Unlike previous bounds on resource conversion (e.g., coherence to entanglement~\cite{CtoE} or entanglement monogamy), our inequality provides a tight bound on the coexistence of local coherence and global mutual information, unifying classical and quantum correlations within a single framework. When considering the conversion among different types of resources, our inequality provides a reliable quantitative relation that unifies classical and quantum correlations within a single framework.


Due to experimental constraints, we only studied a bipartite system. Similar conclusion can also be obtained for larger systems. For instance, we can generalize Eq.~(\ref{trade}) to $N$-partite systems \begin{equation} \label{tradeN}
\sum_{i=1}^N C_{\max}(\rho_{i}) + I(\rho) \leq \sum_{i=1}^N \log_2 d_i,
\end{equation}
where $\rho$ is an $N$-partite quantum state, $\rho_{i}$ is the $i$-th reduced density matrix obtained by tracing out all subsystems except the $i$-th subsystem, and $I(\rho)$ is the QMI for multipartite systems defined as $I(\rho) := \sum_{i=1}^N S(\rho_i) - S(\rho)$. For a detailed proof of Eq.~(\ref{tradeN}), please refer to the supplemental material.
Experimentally, we can apply path degrees of freedom or angular momentum degrees of freedom to increase the system's dimension. We can also use the cascade SPDC effect to generate multibody photonic systems. It is foreseeable that the optical path will become very complex in multibody high-dimensional systems, so designing a reasonable optical path will be a key point in future work.


We thank S. Camalet and Qiongyi He for discussions. This work is supported by the National Natural Science Foundation of China (Grants No.~62475127),  Zhejiang Provincial Natural Science Foundation of China (Grant No. LZ25A040006), and K.C. Wong Magna Fund in Ningbo University.

\section*{APPENDIX}

\subsection{Measure of coherence and maximum coherence}
The relative entropy of coherence defined as
\begin{equation} \label{sm1}
C\left(\varrho\right)=\underset{\delta\in \mathcal{I}}{\min}S\left(\varrho\lVert\delta\right),
\end{equation}
where $\delta$ is a incoherent state, $\mathcal{I}$ is the set of all incoherent states. The meaning of this measure is to find the incoherent state with the smallest relative entropy to the state to be measured from all incoherent states as the quantification of coherence. This incoherent state happens to be the decoherence of the measured state. Thus
\begin{equation} \label{sm2}
C(\varrho)= S(\Delta[\varrho])-S(\varrho).
\end{equation}
This coherence measure satisfies BCP framewokr\cite{coh}. We further introduce maximum coherence instead of coherence for being basis-independent:
\begin{equation} \label{sm3}
C_{\max}(\varrho)=\big[\max_{\{|i\rangle\}} S(\Delta[\varrho])\big]-S(\varrho).
\end{equation}
Here $\max$ is to select a set of bases among all possible bases to maximize $S(\Delta[\varrho])$. After calculation, this formula can be expressed as 
\begin{equation} \label{sm4}
C_{\max}(\varrho)=\log_2 d-S(\varrho),
\end{equation}
where $d$ is the dimension of Hilbert space of $\varrho$ \cite{mco1,mco2}.

\subsection{Derivation of Eq.~(1) in the main text}
Eq.($1$) in the main text is the tradeoff inequality we want to verify in the experiment:
\begin{equation} \label{sm5}
C_{\max}(\varrho_{A})+C_{\max}(\varrho_{B})+I(\varrho_{AB})\leq \log_2{d_A d_B}.
\end{equation}
According to the definition of maximum coherence in Eq.(\ref{sm4}), we have
\begin{eqnarray} 
C_{\max}(\varrho_A)=\log_2 d_A-S(\varrho_A),\label{sm6}\\  
C_{\max}(\varrho_B)=\log_2 d_B-S(\varrho_B).\label{sm62}
\end{eqnarray}
The quantum mutual information(QMI) of $\varrho_{AB}$ is
\begin{equation} \label{sm7}
I(\varrho_{AB})=S(\varrho_{A})+S(\varrho_{B})-S(\varrho_{AB}).
\end{equation}
Put the results of Eqs.(\ref{sm6}-\ref{sm7}) into Eq.(\ref{sm5}) and calculate:
\begin{eqnarray} 
C_{\max}(\varrho_{A})+C_{\max}(\varrho_{B})+I(\varrho_{AB})&=&\log_2{d_A d_B}-S(\varrho_{AB})\nonumber\\
&\leq& \log_2{d_A d_B}.\label{sm8}
\end{eqnarray}
From the above formula we can see that the condition for the equality to hold is that $S(\varrho_{AB})=0$, that is, the state $\varrho_{AB}$ is a pure state.

\subsection{Derivation of Eq.~(4) in the main text}
Eq.~($4$) in the main text is the tradeoff inequality for $N$-partite systems,
\begin{equation} \label{sm5S}
\sum_{i=1}^N C_{\max}(\varrho_{i})+I(\varrho)\leq \sum_{i=1}^N\log_2(d_i),
\end{equation}
where $\varrho$ is an  $N$-partite quantum state, $\varrho_{i}$ is the $i$-th reduced density matrix by tracing out all the subsystems except the $i$-th subsystem.
According to the definition of maximum coherence, we have
\begin{equation} \label{sm6S}
C_{\max}(\varrho_i)=\log_2 d_i-S(\varrho_i),
\end{equation}
The quantum mutual information(QMI) of $\varrho$ in multipartite systems is
\begin{equation} \label{sm7S}
I(\varrho):=\sum_{i=1}^N S(\varrho_i)-S(\varrho).
\end{equation}
Put the results of Eq.(\ref{sm6S}) and Eq.(\ref{sm7S}) into Eq.(\ref{sm5S}) and we can obtain
\begin{equation} \label{sm8S}
\sum_{i=1}^N C_{\max}(\varrho_{i})+I(\varrho)=\sum_{i=1}^N\log_2(d_i)-S(\varrho)\leq \sum_{i=1}^N\log_2(d_i).
\end{equation}
From the above formula we can see that the condition for the equality to hold is that $S(\varrho)=0$, that is, the state $\varrho$ is a pure state.

\end{document}